УДК 534-13

**Description of sound as a self-consistent field in continuous media, analogous to a superconducting state. Theoretical explanation of the experimental Fletcher-Munson curves.**

Braginsky A. Ya.

Southern Federal University

Rostov-na-Donu, Russia

e-mail: a.braginsky@mail.ru,

**Abstract**

We introduce a description of sound waves using the phonon field equivalent to a 4 dimensional second-rank tensor of distortion similar to electromagnetic waves, which are described by a 4-vector of the electromagnetic field. The exact wave solutions of the state equations of the Kadić-Edelen Lagrangian are obtained for the phonon field in a continuous isotropic medium. We demonstrate that the state of the medium with the phonon field is analogous to the self-consistent superconducting state of the electromagnetic field described by the London equation. It is shown that the energy density of the phonon field in the air is proportional both to the quares of pressure and of frequency, which corresponds to the experimental Fletcher-Munson curves.

**ВВЕДЕНИЕ**

Как известно [1], плотность энергии звуковой волны, полученной из уравнений Эйлера, задается выражением:

$$E = \beta p^2, \qquad (1)$$

где $p$ - давление, $\beta$ - коэффициент сжимаемости. Однако, плотность энергии (1) не соответствует фоновой диаграмме Флетчера-Мэнсона (рис. 1) [2,3], поскольку в (1) отсутствует явная зависимость энергии от частоты.

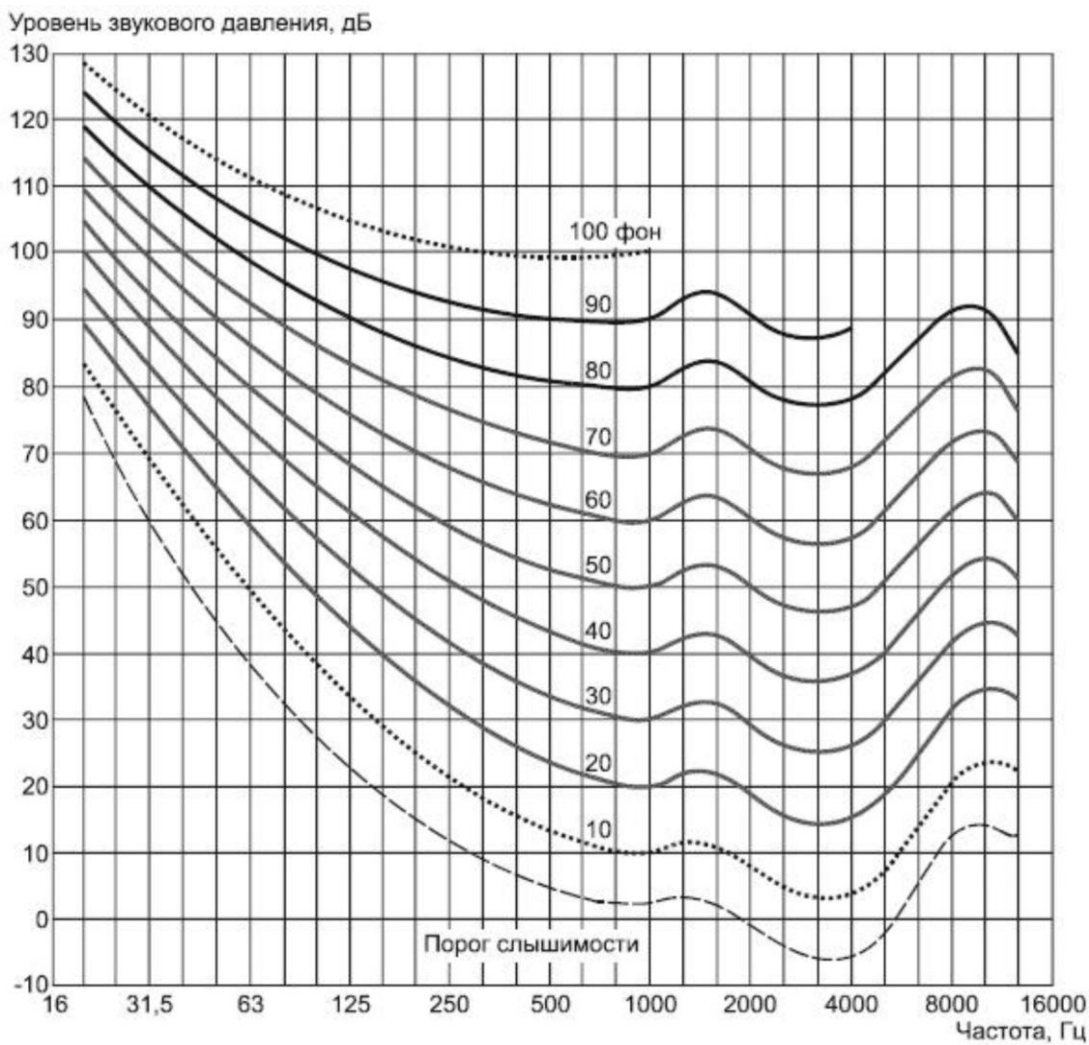

Рис. 1. Диаграмма Флетчера-Мэнсона [ГОСТ Р ИСО 226-2009].

На рис. 1 изображены линии – изофоны, которые соответствуют одинаковой интенсивности (громкости) звука, в зависимости от давления звуковой волны и частоты. Зависимость интенсивности звука от частоты столь очевидна и велика, что на диаграмме рис. 1 для давления используется логарифмическая шкала децибелов. Интенсивность или громкость звука с увеличением частоты при постоянном давлении изменяется более чем в миллион раз (60 дБ). При этом теоретическое обоснование диаграммы Флетчера-Менсона отсутствует, так как считается, что она субъективная.

Частотную зависимость диаграммы Флетчера-Менсона вряд ли можно объяснить только субъективными особенностями слуха. Представляется, что предложение измерения интенсивности (громкости) звука в фонах было сделано для того, чтобы в описании диаграммы рис. 1 отсутствовало понятие плотности энергии звуковой волны (1), так как механическая плотность энергии (1) не описывает диаграмму Флетчера-Менсона.

Вообще говоря, описание звука как механических колебаний сплошной среды не очевидно. Есть много механических колебаний, которые не сопровождаются звуком. В данной работе предложено описывать звук по аналогии с описанием электромагнитных волн. В качестве основной переменной для описания звука используется 4-тензор дисторсии или **фононный потенциал**: $A_{pj}$, $\upsilon_p$, который является обобщением пространственных производных вектора смещения $\partial u_i/\partial X_j$, $\partial u_i/\partial T$. В работе показано, что плотность энергии фононного поля описывает диаграмму Флетчера-Менсона рис. 1.

Тензор дисторсии $A_{pj}$ как обобщение $\partial u_i / \partial X_j$ был введен в теории упругости [4] для описания плотности дислокаций:

$$\rho_{ij} = -e_{jkn} \partial A_{in} / \partial X_k, \qquad (2)$$

когда вектор смещений $u_i$ испытывает разрыв и не определен. При этом считается, что $A_{pj}$ является непрерывной дифференцируемой функцией. Обобщение $\partial u_i / \partial T$ в виде поля скорости $\upsilon_p$, как четвертой компоненты тензора дисторсии, было сделано в калибровочной теории дислокаций Кадича-Эделена [5,6] (п. 4).

Идея использовать тензор дисторсии в качестве фононного потенциала появилась не случайно. Он выполняет роль компенсирующего поля параметра порядка (ПП) в неоднородной теории Ландау, аналогичную вектору электромагнитного потенциала в теории Гинзбурга-Ландау [7] (п. 2). Как известно, в калибровочной теории компенсирующее поле содержится в неравновесном потенциале не только в удлиненной производной ПП, но и в виде инвариантных антисимметричных комбинаций производных, например (2). В п. 6 статьи показано, что учет в лагранжиане фононного поля инвариантов, составленных из производных 4-тензора дисторсии, аналогичных свертке компонент тензора электромагнитного поля $-\frac{1}{16\pi} F_{\mu\nu} F^{\mu\nu}$ [8], приводит к плотности энергии для продольной звуковой волны в воздухе в виде:

$$E = 4\beta p^2 \frac{\omega^2}{\omega_0^2}, \qquad (3)$$

где $\omega_0$ - минимальная частота фононного поля, см. (42). Квадратичная зависимость плотности энергии от частоты в (3), для волновых решений $A_{pj}$, $\upsilon_p$, следует из квадратичного лагранжиана фононного поля, аналогичного лагранжиану электромагнитного поля. Квадратичная зависимость плотности энергии от давления в (3) следует из линейной зависимости между сопряженными величинами в сплошной изотропной среде: $\sigma_{pj} = \beta^{-1} A_{pj}$, аналогичной закону Гука [4], и определения тензора напряжений в воздухе $\sigma_{pj} = \delta_{pj} p$.

Плотность энергии (3) соответствует фоновой диаграмме Флетчера-Мэнсона рис.1. Согласно (3) при постоянной плотности энергии и изменении частоты в звуковом диапазоне от 20Гц до 20КГц - давление уменьшается в миллион раз. В п.6 построены линии равной плотности энергии в логарифмических координатах частота-давление и дан сравнительный анализ с субъективной диаграммой Флетчера-Менсона.

Существующая теория звуковых волн в воздухе [1] базируется не на принципе наименьшего действия, а на уравнениях гидродинамики Эйлера:

$$\rho \frac{\partial \vec{v}}{\partial T} = -\frac{1}{2} \rho \vec{\nabla}(\vec{v}^2) + \rho [\vec{v} \times (\vec{\nabla} \times \vec{v})] - \vec{\nabla} p, \qquad (4)$$

где $\rho$ - плотность сплошной среды, $\vec{v} = \partial \vec{X} / \partial T$ - вектор скорости. Как известно [1], для описания звука используют линеаризованные уравнения Эйлера, которые получаются, если аннулировать в (7) нелинейные члены (первые два слагаемых в правой части). Такая ревизия уравнений Эйлера обосновывается малой амплитудой звуковых колебаний. Однако при этом не учитывается, что градиент скорости может быть большим, например, при высоких частотах. Более того, полученные решения для амплитуд давления и скорости прямо пропорциональны частоте [1]. Поэтому нелинейными членами в (4) нельзя пренебрегать при описании звука. Очевидно, что любые нелинейности приводят к тому, что частота звука будет меняться при изменении амплитуды. На практике этого не наблюдается. Поэтому есть основания считать, что звук изначально описывается линейными дифференциальными уравнениями. Вряд ли, эти уравнения являются линеаризованными уравнениями Эйлера. Тем более, что механическое описание звука (1) [1] не объясняет фоновую диаграмму Флетчера-Менсона рис.1.

В п.5 получены линейные уравнения состояния для фононного поля, аналогичные уравнениям Максвелла. Показано, что в сплошной изотропной среде уравнения состояния имеют точные волновые решения, соответствующие звуку, которые описывают (3) диаграмму Флетчера-Мэнсона. При этом фононное поле, как компенсирующее поле ПП, в сплошной среде обладает свойствами подобными массивному бозону Хиггса [9,10].

**КОМПЕНСИРУЮЩЕЕ ПОЛЕ ДИСТОРСИИ**

Предложение введения новых переменных 4-тензора дисторсии $A_{pj}$, $\upsilon_p$ для описания звука, по аналогии с 4-вектором электромагнитного поля $A_j$, $\varphi$, появилось не случайно. Оно возникло при рассмотрении неоднородных потенциалов с удлиненной производной ПП в теории фазовых переходов Ландау. В этих потенциалах тензор дисторсии $A_{pj}$ выполнял роль компенсирующего поля ПП, аналогичную роли электромагнитного потенциала $A_j$ в полевой теории электродинамики [7]. Приведем в качестве примера несколько моделей, где использовался тензор дисторсии в качестве компенсирующего поля.

Необходимость введения тензора дисторсии $A_{pj}$ как компенсирующего поля в теории Ландау возникла при исследовании состояний с локальными трансляционными свойствами ПП, когда вектор $\vec{k} = \vec{k}(\vec{X})$ [11]. В этом случае, под действием оператора элементарной трансляции $a_j$ на производную ПП, появляется дополнительное слагаемое, связанное с зависимостью $\vec{k} = \vec{k}(\vec{X})$:

$$\hat{a}_q \frac{\partial \eta}{\partial X_j} = \frac{\partial}{\partial X_j}\left(e^{ik_q(\vec{X})a_q}\eta(\vec{X})\right) = e^{ik_q(\vec{X})a_q}\left[i\frac{\partial k_q(\vec{X})a_q}{\partial X_j}\eta(\vec{X}) + \frac{\partial \eta(\vec{X})}{\partial X_j}\right].$$ Чтобы

аннулировать первый член в квадратных скобках, и тем самым диагонализировать базис компонент ПП и его производных, удлиняют производную ПП:

$$D_j \eta = \left(\frac{\partial}{\partial X_j} - i\sum_p \kappa_p A_{pj}\right)\eta \qquad (5)$$

с помощью тензора $A_{pj}$, который преобразуется следующим образом при элементарных трансляциях:

$$\hat{a}_q(\kappa_p A_{pj}) = \kappa_p A_{pj} + \delta_{pq}\frac{\partial(k_p(\vec{X})a_q)}{\partial X_j}. \qquad (6)$$

Здесь $\kappa_p$ - феноменологический заряд, имеющий размерность волнового вектора, а $\delta_{qp}$ - символ Кронекера.

Наиболее известная, модель, в которой необходимо использовать тензор дисторсии $A_{pj}$ в качестве компенсирующего поля - это модель де Жена деформированного SmA [12,13]. Де Жен обратил внимание на то, что деформированный SmA ведет себя подобно сверхпроводникам в магнитном поле, и экранирует деформацию путем образования дислокаций. При этом фазовая диаграмма деформированного SmA была эквивалентна фазовой диаграмме сверхпроводников второго рода [7]. В качестве основных варьируемых переменных де Жен использовал двухкомпонентный ПП $\psi(\vec{X})$, $\psi^*(\vec{X})$ с вектором $\vec{k}$, описывающий переход нематик – SmA, и директор $\vec{n} = \vec{n}(\vec{X})$ - единичный вектор нормали перпендикулярный слоям SmA. Так как вектор $\vec{k} = 2\pi\vec{n}/d$, то зависимость $\vec{n} = \vec{n}(\vec{X})$ означала, что вектор $\vec{k}$ зависит от $\vec{X}$ (здесь $d$ - расстояние между слоями в SmA). Из (5,6) следует, что для построения инвариантного потенциала Ландау в модели де Жена надо удлинить производную смектического ПП путем введения в теорию компенсирующего тензорного поля $A_{pj}$. При этом инвариантные производные компенсирующего тензора (2) будут соответствовать плотности дислокаций, что позволяет интерпретировать компенсирующее поле (5,6) как тензор дисторсии [4,11,14].

При описании сверхпроводящего состояния ПП с $\vec{k} = \vec{k}(\vec{X})$ в модели Г-Л [7,14] также необходимо ввести компенсирующий тензор $A_{pj}$ (6), который входит в удлиненную производную ПП линейным образом:

$$D_j \psi_{\vec{k}} = \left( \frac{\partial}{\partial X_j} - i \sum_p \kappa_p A_{pj} - i \frac{e}{\hbar c_e} A_j \right) \psi_{\vec{k}} . \qquad (7)$$

В (7) минимальное взаимодействие ПП и тензора $A_{pj}$ отвечает за электрон-фононное взаимодействие в феноменологической теории Гинзбурга-Ландау. Это обстоятельство позволяет интерпретировать тензор $A_{pj}$, с трансформационными свойствами (6), как фононный потенциал.

В п.3 доказано, что тензорное поле $A_{pj}$ (6), введенное в (5) с целью компенсировать изменение вектора $\vec{k} = \vec{k}(\vec{X})$ ПП, и тензор дисторсии, введенный в [4] как обобщение $\partial u_i / \partial X_j$ для определения плотности дислокаций (2), совпадают.

Так как компенсирующее поле является независимой переменной, то оно входит в потенциал не только в удлиненных производных, но и в виде градиентных инвариантов (2), как, например, $\vec{B} = rot\vec{A}$ в теории Гинзбурга-Ландау. С другой стороны известно, что электромагнитные волны описываются с помощью уравнений Максвелла, которые могут быть получены из вариации лагранжиана $L = \frac{1}{8\pi}(\vec{E}^2 - \vec{B}^2)$ [8]. Построим аналогичную вариационную теорию для фононного потенциала $A_{pj}$, $\upsilon_p$. Для этого надо рассмотреть динамическую модель, и перейти от неравновесного потенциала Ландау к лагранжиану. Задача построения лагранжиана фононного поля, по сути дела была выполнена в калибровочной теории дислокаций Кадича-Эделена [5,6], которая подробно описана в п.4.

### НАПРЯЖЕННОСТЬ ФОНОННОГО ПОЛЯ

Необходимость перехода от пространственных производных вектора смещения $\partial u_i / \partial X_j$, $\partial u_i / \partial T$ к новым упругим переменным 4-тензора дисторсии $A_{pj}$, $\upsilon_p$ можно проследить при переходе от сплошной среды к кристаллической решетке.

Как известно [4], в теории упругости в качестве основной варьируемой переменной выступает вектор смещения $u_i$, который определяет тензор деформаций. По определению, вектор смещения и его производные являются инвариантами подгруппы трансляций в пространстве $\{\vec{X}\}$. Перейдем от сплошной среды к среде, в которую помещена кристаллическая решетка. Тогда согласно теории Ландау надо потребовать инвариантность тензора $\partial u_i / \partial X_j$ не только по отношению к произвольной трансляции в $\{\vec{X}\}$, но и по отношению к трансляциям на элементарный период кристаллической решетки $a_i$. Этого требования не было, при рассмотрении сплошной среды, так как отсутствовала решетка.

В общем случае, в деформированном состоянии реализуется ситуация с локальной подгруппой элементарных трансляций $a_i = a_i(\vec{X})$. Зависимость $a_i = a_i(\vec{X})$ вытекает из самого понятия неоднородной деформации кристаллической решетки. Здесь предполагается, что в макроскопическом малом объеме с координатой $\vec{X}$ существует период $a_i = a_i(\vec{X})$, согласно положениям неоднородной теории [15].

В [15] Лифшиц впервые показал, что в теории Ландау должны исследоваться неоднородные состояния, и сформулировал вариационный принцип в теории Ландау. Как известно [11,15], неоднородные состояния в теории Лифшица исследуются только в макроскопических координатах. Чтобы подчеркнуть макроскопический характер координат Лифшица, в данной статье координаты обозначаем большими буквами.

Из определения вектора смещения $\vec{u}(\vec{X}) = \vec{X} - \vec{X}_0$, следует, что:

$$\hat{a}_i(u_i) = u_i(\vec{X}) + a_i(\vec{X}) - a_i(\vec{X}_0), \qquad (8)$$

где $a_i(\vec{X}_0)$ - период решетки в недеформированном состоянии, $a_i(\vec{X})$ – период решетки после деформации в точке $\vec{X}$. Согласно (8) производные вектора смещения при трансляции на элементарный период преобразуется следующим образом:

$$\hat{a}_i(\partial u_i/\partial X_j) = \partial u_i/\partial X_j + \partial a_i/\partial X_j, \qquad (9)$$

аналогичные рассуждения приводят к выражению:

$$\hat{a}_i(\partial u_i/\partial T) = \partial u_i/\partial T + \partial a_i/\partial T. \qquad (10)$$

Следовательно, тензор деформации не является инвариантом подгруппы элементарных трансляций всегда, когда учитывается локальная симметрия деформированной решетки $a_i(\vec{X})$ (9,10). При этом не инвариантность по отношению к элементарной трансляции $a_i$ не означает, что тензор перестал быть тензором (инвариантом пространственной подгруппы трансляций в $\{\vec{X}\}$).

Вообще говоря, ситуация, когда в каждом макроскопическом малом объеме решетка обладает локальной трансляционной симметрией, имеет место для абсолютно хрупких решеток, таких как жидкие кристаллы [12,13] или гелимагнетики [16-18]. Вихри Абрикосова [7] также можно представить как дислокации подрешетки сверхпроводящих электронов, плотность которых в вихре равна нулю. Это следует из того, что линии плотности дислокаций и магнитного поля в смешанном сверхпроводящем состоянии совпадают, так как они задаются одной удлиненной производной (7).

Случай $a_i = a_i(\vec{X})$ относится к такой модели деформации, когда решетка рвется в каждой точке $\vec{X}$ с образованием дислокаций. Действительно, рассмотрим двумерную модель, в которой период изменяется в перпендикулярном направлении $a_1 = a_1(X_2)$. В этом случае при совмещении соседних областей с различными периодами образуются краевые дислокаций, линии которых перпендикулярны двумерной плоскости [18]. Откуда следует, что в случае локальной трансляционной симметрии деформированной решетки $a_i = a_i(\vec{X})$, ее упругие свойства должны описываться тензором дисторсии $A_{pj}$ [4], который задает плотность дислокаций (2).

Согласно (9) трансформационные свойства тензора дисторсии $A_{pj}$, который обобщает $\partial u_i/\partial X_j$, имеют вид:

$$\hat{a}_q(A_{pj}) = A_{pj} + \delta_{qp}\,\partial a_q/\partial X_j. \qquad (11)$$

Из (10), получаем трансформационные свойства для $\upsilon_p$, как обобщения $\partial u_i/\partial T$:

$$\hat{a}_q(\upsilon_p) = \upsilon_p + \delta_{qp}\partial a_q/\partial T. \qquad (12)$$

Покажем, что трансформационные свойства компенсирующего поля $A_{pj}$ в (6) соответствуют (11). Положим что в (6) локальна группа $a_p = a_p(\vec{X})$, а вектор $k_p$ не зависит от $\vec{X}$ и задает период низко-симметричной фазы: $k_p a_p = 2\pi$. Тогда выражение (6) переходит в (11) когда феноменологический заряд в (5) равен вектору $k_p$: $\kappa_p = k_p$. Как известно [7,14], феноменологический заряд в удлиненной производной задает квант потока вихря компенсирующего поля (в теории сверхпроводимости [7] это квант магнитного потока). Для тензора $\rho_{pj}$ (2) поток $\int_{S_L} \rho_{pj} ds_j$ равен вектору Бюргерса $B_p$, из определения тензора дисторсии:
$B_p = -\oint_L A_{pj} dx_j$ [4]. Следовательно, минимальный вектор Бюргерса связан с феноменологическим зарядом (5,7) соотношением: $B_{p\min} = \dfrac{2\pi}{\kappa_p}$ (данный вывод полностью аналогичен выводу выражения для кванта магнитного потока [7,14]). С другой стороны, минимальный вектор Бюргерса равен периоду решетки $a_p = \dfrac{2\pi}{\kappa_p}$, откуда получаем $\kappa_p = k_p$. Следовательно, компенсирующее поле $A_{pj}$ в (6) это тензор дисторсии (2), который является обобщением $\partial u_i/\partial X_j$ (11) для состояний с дислокациями.

Таким образом, при описании деформированного состояния кристалла с локальной трансляционной симметрией $a_i = a_i(\vec{X})$, от трех независимых степеней свободы вектора смещения $u_i$ надо перейти к двенадцати степеням свободы $A_{pj}$, $\upsilon_p$, связанным между собой трансформационными условиями (11,12).

При формулировке вариационной задачи для фононного поля $A_{pj}$, $\upsilon_p$ в лагранжиане надо учесть его пространственные и временные производные. Согласно (11,12) инвариантные комбинации производных фононного поля имеют вид:
$$\rho_{pj} = -e_{jkn} \partial A_{pn}/\partial X_k, \tag{2}$$
$$\varepsilon_{pj} = -\partial \upsilon_p/\partial X_j + \partial A_{pj}/\partial T. \tag{13}$$

Сигнатура в (2) выбрана в соответствии с прямыми вычислениями, проделанными в [4]. Сигнатура выражения (13) выбрана исходя из сигнатуры уравнения Бернулли: $v^2/2 + p/\rho = const$. В стационарном случае напряженность $\varepsilon_{pj}$ (13) приводит к силе $f_i = -p_p \partial \upsilon_p/\partial X_i$ эквивалентной потенциальной силе $\vec{f} = -\dfrac{1}{2}\rho\vec{\nabla}(\vec{v}^2)$ в уравнении Эйлера (4), которая дает вклад $v^2/2$ в уравнение Бернулли. Точный вывод выражений (2,13) можно получить из минимума действия для движения частицы с импульсом $p_p$ в фононном поле $A_{pj}$, $\upsilon_p$, по аналогии с выводом сил, действующих на электрический заряд в электромагнитном поле [8,14].

Центрально-симметричная сила, действующая на импульс в фононном поле, имеет вид:
$$f_i^C = p_p(-\partial \upsilon_p/\partial X_i + \partial A_{pi}/\partial T), \tag{14}$$
а антисимметричная сила, действующая на поток импульса, имеет вид: $f_i^A = e_{ikn} p_p v_k \rho_{pn}$. Так как тензор напряжений согласно уравнению Ньютона равен потоку импульса $\sigma_{pk} = -p_p v_k$ (см. 26), то антисимметричная сила $f_i^A$ - это известная сила Пича-Келлера [4,19]:
$$f_i^A = -e_{ikn}\sigma_{pk}\rho_{pn}, \tag{15}$$
аналогичная силе Лоренца $f_i^L = e_{ikn} j_k B_n$.

Величины $\varepsilon_{pj}$ и $\rho_{pj}$ (2,13), инвариантные при преобразованиях (11,12), являются наблюдаемыми характеристиками фононного поля, аналогичными электрической напряженности $E_j$ и магнитной индукции $B_j$ в электродинамике. Если про непрерывную функцию $\rho_{pj}$ известно, что она связана с дислокациями [4], то про напряженность $\varepsilon_{pj}$, которая соответствует электрической напряженности $E_j = -\partial \varphi / \partial X_j - \partial A_j / \partial T$ в электродинамике, практически ничего не известно. Приведем соображения, которые позволяют связать компоненты $\varepsilon_{pj}$ со звуком.

Обратим внимание на то, что градиент скорости $\partial \upsilon_p / \partial X_j$ связан с формой духовых инструментов, которые имеют или расширения на конце, или отверстия. При выдувании воздуха из них, образуется звук в виде градиента скорости. Об этой особенности духовых инструментов известно давно. Производную $\partial A_{pj} / \partial T$ в воздухе можно связать с изменением давления. Для этого обобщим закон Гука на тензор дисторсии, тогда для изотропного случая получим: $\sigma_{pj} = \beta^{-1} A_{pj}$. Так как в воздухе $\sigma_{pj} = -\delta_{pj} p$, то $A_{pj} = -\delta_{pj} \beta p$. Следовательно, в однородном случае, напряженность $\varepsilon_{pj}$ пропорциональна скорости давления $\varepsilon_{pj} = -\delta_{pj} \beta \dfrac{\partial p}{\partial T}$. Изменение давления со временем это, по сути дела, то, что мы слышим. Поэтому, по аналогии с электрической напряженностью $E_j$, $\varepsilon_{pj}$ будем называть акустической напряженностью.

Заметим, что напряженности фононного поля (2,13) описывают такие степени свободы, которые не зависят от градиентных преобразований: $A_{ij} \to A_{ij} + \partial u_i / \partial X_j$, $\upsilon_i \to \upsilon_i + \partial u_i / \partial T$. Поэтому $\varepsilon_{pj}$ и $\rho_{pj}$ - это дополнительные степени свободы фононного поля, которые не являются механическими (они не зависят от вектора смещения $u_i$). Согласно (2,13), они связаны между собой условиями, аналогичными первой паре уравнений Максвелла:

$$\frac{\partial \rho_{pj}}{\partial X_j} = 0, \quad \frac{\partial \rho_{pj}}{\partial T} = -e_{jmn} \frac{\partial \varepsilon_{pn}}{\partial X_m}.$$

## ЛАГРАНЖИАН ФОНОННОГО ПОЛЯ

Впервые 4-тензор дисторсии $A'_{pj}$, $\upsilon'_p$ ввели Кадич с Эделеном в калибровочной теории дислокаций [5]. Здесь в обозначении мы поставили штрих, чтобы отличать 4-тензор дисторсии Кадича-Эделена от введенного выше $A_{pj}$, $\upsilon_p$ - фононного потенциала (11,12), полученного как обобщение $\partial u_i / \partial X_j$, $\partial u_i / \partial T$. В [6] дана современная трактовка теории Кадича-Эделена и построен лагранжиан для 4-тензора дисторсии $A'_{pj}$, $\upsilon'_p$. Покажем, что 4-тензор дисторсии Кадича-Эделена связан с фононным потенциалом соотношением: $A'_{ij} = -A_{ij}$, $\upsilon'_i = -\upsilon_i$.

Изначально теория Кадича-Эделена возникла из линейного приближения теории Янга-Миллса [5]. Современное ее изложение [6] исходит из постулата о локальности пространственной подгруппы трансляций $b_i = b_i(X_i, T)$ в пространстве $\{\vec{X}\}$, в котором принимает свои значения вектор смещения:

$$\hat{b}_i(u_i) = u_i(\vec{X}, T) + b_i(\vec{X}, T). \tag{16}$$

В этом случае пространственные и временные производные от вектора смещения $\partial u_i / \partial X_j$, $\partial u_i / \partial T$ становятся не инвариантными по отношению к локальной подгруппе трансляций:

$$\hat{b}_i(\partial u_i/\partial X_j) = \partial u_i/\partial X_j + \partial b_i/\partial X_j,$$
$$\hat{b}_i(\partial u_i/\partial T) = \partial u_i/\partial T + \partial b_i/\partial T.$$

Для того, чтобы построить функцию Лагранжа инвариантную при преобразованиях (16) вводят в теорию дополнительные компенсирующие поля дисторсии $A'_{ij}$ и скорости $\upsilon'_i$, которые преобразуются следующим образом:

$$\hat{b}_i(A'_{ij}) = A'_{ij} - \partial b_i/\partial X_j, \quad \hat{b}_i(\upsilon'_i) = \upsilon'_i - \partial b_i/\partial T \tag{17}$$

Тогда, заменяя производные $\partial u_i/\partial X_j$, $\partial u_i/\partial T$ на трансляционно-инвариантные комбинации:

$$D_j u_i = \partial u_i/\partial X_j + A'_{ij}, \quad D_0 u_i = \partial u_i/\partial T + \upsilon'_i \tag{18}$$

и учитывая собственные инварианты тензора дисторсии и вектора скорости, составленные из производных:

$$\rho_{ij} = e_{jkn} \partial A'_{in}/\partial X_k, \tag{19}$$
$$\varepsilon_{ij} = \partial \upsilon'_i/X_j - \partial A'_{ij}/\partial T, \tag{20}$$

получают трансляционно-инвариантную функцию Лагранжа в виде:

$$L(X_i, T) = L(D_0 u_i, D_j u_i, \rho_{ij}, \varepsilon_{ij}), \tag{21}$$

где под $\rho_{ij}$ - понимают плотность дислокаций [4-6], а $\varepsilon_{ij}$ - градиентный инвариант, аналогичный вектору напряженности в электродинамике. При этом считается, что функция Лагранжа (21), зависящая от удлиненных производных вектора смещения, является квадратичной формой:

$$L(D_0 u_i, D_j u_i) = \mathrm{K}(D_0 u_i) - F(D_j u_i). \tag{22}$$

Она равна разности между кинетической и потенциальной упругой энергией.

Потенциальная энергия $F(D_j u_i)$ представляет собой потенциал теории упругости [4], в котором произвели замену $\partial u_i/\partial X_j$ на $D_j u_i$ (18). Кинетическая энергия $\mathrm{K}(D_0 u_i)$ представляет собой квадратичную функцию скорости, где заменили $\partial u_i/\partial T$ на $D_0 u_i$ (18). Зависимость лагранжиана (21) от $\rho_{ij}$, $\varepsilon_{ij}$ была получена по аналогии со сверткой тензора электромагнитного поля в электродинамике $-\frac{1}{16\pi}F_{\mu\nu}F^{\mu\nu}$:

$$L = \frac{\gamma}{2}(\frac{1}{c^2}\varepsilon_{pj}\varepsilon_{pj} - \rho_{pj}\rho_{pj}), \tag{23}$$

здесь $\gamma$ - размерный множитель, $c$ - скорость звука.
Из (18,22) следует, что:
$$\partial L/\partial A'_{ij} = -\sigma_{ij}, \; \partial L/\partial \upsilon'_i = p_i, \tag{24}$$

где $\sigma_{ij}$ - тензор напряжений, а $p_i$ - вектор импульса. Откуда, без конкретизации вида зависимости (22), можно записать лагранжиан в виде:

$$L = p_i \upsilon'_i - \sigma_{ij} A'_{ij} + \frac{\gamma}{2}(\frac{1}{c^2}\varepsilon_{ij}\varepsilon_{ij} - \rho_{ij}\rho_{ij}), \tag{25}$$

где $p_i$ и $\sigma_{ij}$ - внешние параметры. Из уравнений состояния $\delta L/\delta A'_{ij} = 0$, $\delta L/\delta \upsilon'_i = 0$ для лагранжиана (25), легко получить уравнения второго закона Ньютона в виде дифференциальной записи закона сохранения импульса:

$$\partial \sigma_{ij}/\partial X_j = \partial p_i/\partial T. \tag{26}$$

Для этого достаточно на уравнение $\delta L/\delta A'_{ij} = 0$ подействовать оператором дивергенции по второму индексу, а уравнение $\delta L/\delta \upsilon'_i = 0$ продифференцировать по времени, и сложить полученные выражения.

В [5,6] показано, что 4-тензор дисторсии $A'_{pj}$, $\upsilon'_p$ удовлетворяет псевдолоренцевому калибровочному условию:

$$\frac{\partial A'_{pj}}{X_j} - \frac{1}{c^2}\frac{\partial \upsilon'_p}{\partial T} = 0 \,. \tag{27}$$

Сигнатура (27) связана с трансформационными свойствами (17).

Как следует из п. 3, основной постулат теории Кадича-Эделена, о локальности подгруппы трансляций (16) имеет обоснование. Он справедлив для состояний с локальной трансляционной симметрией решетки (8-10), с оговоркой, что локальна не подгруппа трансляций в $\{\vec{X}\}$, а трансляционная симметрия решетки $a_i(\vec{X})$. Однако выводы, которые были сделаны в п. 3 и в теории Кадича-Эделена – различные. В п. 3 заменили $\partial u_i/\partial X_j$ на $A_{ij}$, чтобы определить плотность дислокаций (2), а сам тензор дисторсии было предложено использовать в качестве компенсирующего поля для ПП с локальными трансляционными свойствами (5,7) п. 2. В теории Кадича-Эделена оставили вектор смещения в качестве основной варьируемой переменной и ввели инвариантные комбинации (18,22).

Согласно трансформационным свойствам (9-12) и (17) 4-тензор дисторсии Кадича-Эделена связан с фононным потенциалом соотношениями: $A'_{ij} = -A_{ij}$, $\upsilon'_i = -\upsilon_i$. Легко увидеть, что в этом случае выражения для напряженностей (2,13) и (19,20) – совпадают, а калибровочное условие (27) для фононного поля имеет вид:

$$\frac{\partial A_{pj}}{X_j} - \frac{1}{c^2}\frac{\partial \upsilon_p}{\partial T} = 0 \,. \tag{28}$$

Сопряжения (24), согласно (18,22) и определению фононного поля, как обобщения $\partial u_i/\partial X_j$, $\partial u_i/\partial T$, остаются без изменений:

$$\partial L/\partial A_{ij} = -\sigma_{ij}, \ \partial L/\partial \upsilon_i = p_i \,. \tag{29}$$

Следовательно, лагранжиан фононного поля имеет вид:

$$L = p_i\upsilon_i - \sigma_{ij}A_{ij} + \frac{\gamma}{2}(\frac{1}{c^2}\varepsilon_{ij}\varepsilon_{ij} - \rho_{ij}\rho_{ij}) \tag{30}$$

Калибровочное условие (28) для фононного поля в кристаллической решетке имеет простой физический смысл. Так как выражение (28) должно быть инвариантно при действии оператора элементарной трансляции (11,12), то $a_p$ удовлетворяет уравнению Даламбера:

$$(\Delta - \frac{1}{c^2}\frac{\partial^2}{\partial T^2})a_p = 0 \,, \tag{31}$$

Калибровочное условие (28) фиксирует состояние решетки, при котором исследуется фононное поле. Условие (31) соответствует состоянию, в котором по решетке распространяется волна $a_p$ со скоростью звука $c$. Это объясняет, почему звук сопровождается механическими колебаниями решетки.

По поводу теории Кадича-Эделена заметим, что в ней нет понятия свободного поля дисторсии. Тензор дисторсии и вектор смещения не могут существовать друг без друга по определению (18). Это происходит потому, что комбинации (18) не являются удлиненными производными, они не аннулируются при $u_i = 0$, в отличие от удлиненных производных (5,7) когда ПП равен нулю. Выражения (18) и (22) перестают быть трансляционно-инвариантными комбинациями при $u_i = 0$ (17), поэтому в [5,6] не исследуется свободное поле 4-тензора дисторсии.

По сути дела, в теории Кадича-Эделена исследуется разность $\partial u_i / \partial X_j - A_{ij}$ (18) между производными смещения и их обобщением (2) [4], когда вектор смещения не является наблюдаемой величиной (16). Вряд ли такая теория может привести к интересным физическим результатам.

### УРАВНЕНИЯ СОСТОЯНИЯ ФОНОННОГО ПОЛЯ В СПЛОШНОЙ ИЗОТРОПОНОЙ СРЕДЕ

Исследуем фононное поле $A_{pj}$, $\upsilon_p$ в сплошной среде. Перейдем от кристаллической решетки, исследуемой в п. 3, к сплошной среде. В этом случае фононный потенциал уже не будет отвечать условиям (11,12), так как нет дополнительной симметрии, задаваемой подгруппой элементарных трансляций $a_p$. С другой стороны, компоненты фононного потенциала $A_{pj}$, $\upsilon_p$ являются степенями свободы, которые описывают физическое состояние сплошной среды. Они, например, определяют плотность дислокаций в сплошной среде (2) [4].

Отличие в определении плотности дислокаций в сплошной среде, от плотности дислокаций в модели с решеткой, заключается в том, что в сплошной среде вектор Бюргерса не является дискретным. Действительно, в сплошной среде нет вектора $k_p$, который сопряжен элементарной трансляции $a_p$ кристаллической решетки. Следовательно, в сплошной среде не может быть и минимального взаимодействия между фононным полем и ПП с феноменологическим зарядом $\kappa_p$ (5), которое задает минимальный вектор Бюргерса $B_{p\min} = \frac{2\pi}{\kappa_p}$. Как только возникает минимальное взаимодействие, то появляется и подрешетка, которую описывает ПП с вектором $k_p$.

Таким образом, компоненты фононного поля - вектор скорости $\upsilon_p$ и тензор дисторсии $A_{pj}$, являются наблюдаемыми величинами в отсутствии трансформационных условий (11,12), так как в приближении сплошной среды $a_p(\vec{X}) = 0$ в (11,12).

Лагранжиана (30) достаточно чтобы получить уравнения состояния для фононного поля, аналогичные второй паре уравнений Максвелла [8]. Согласно (2,13) уравнения состояния $\delta L / \delta A_{pj} = 0$, $\delta L / \delta \upsilon_p = 0$ для лагранжиана (30) имеют вид:

$$\sigma_{pj} = \gamma \, e_{jki} \frac{\partial \rho_{pi}}{\partial X_k} - \frac{\gamma}{c^2} \frac{\partial \varepsilon_{pj}}{\partial T}. \tag{32}$$

$$p_p = -\frac{\gamma}{c^2} \frac{\partial \varepsilon_{pj}}{\partial X_j}. \tag{33}$$

Здесь тензор напряжения и вектор импульса не определяются уравнениями состояния, как например плотность тока в модели Гинзбурга-Ландау [7], а являются внешними параметрами.

При условии (28) выражения (32,33) трансформируются в выражения:

$$\frac{1}{\gamma} \sigma_{pj} = (\Delta - \frac{1}{c^2} \frac{\partial^2}{\partial T^2}) A_{pj}. \tag{34}$$

$$\frac{c^2}{\gamma} p_p = (\Delta - \frac{1}{c^2} \frac{\partial^2}{\partial T^2}) \upsilon_p. \tag{35}$$

В данной работе фононное поле исследуется в сплошной среде, которая характеризуется плотностью $\rho$. Очевидно, что при наличии поля скорости $\upsilon_i$ импульс будет связан с ним соотношением: $p_i = \rho \upsilon_i$. Поэтому, при исследовании волн фононного поля в сплошной среде, импульс в (35) нельзя аннулировать, как, например, аннулируется заряд в уравнениях Максвелла при описании электромагнитных волн [8]. Обобщая закон Гука, для сплошной изотропной среды, положим:

$$\sigma_{pj} = \beta^{-1} A_{pj}, \qquad (36)$$

где $\beta^{-1} = K$ - модуль упругости. Тогда уравнения (34,35) принимают вид:

$$\frac{1}{\gamma \beta} A_{pj} = (\Delta - \frac{1}{c^2} \frac{\partial^2}{\partial T^2}) A_{pj}. \qquad (37)$$

$$\frac{c^2 \rho}{\gamma} \upsilon_p = (\Delta - \frac{1}{c^2} \frac{\partial^2}{\partial T^2}) \upsilon_p. \qquad (38)$$

При условии $\beta^{-1} = c^2 \rho$ уравнения (37,38) эквивалентны, и их решения обладают одинаковым спектром. Как известно [1,4], выражение $c^2 = \beta^{-1} \rho^{-1}$ определяет квадрат скорости звука:

$$c = \sqrt{1/\beta \rho}. \qquad (39)$$

Выражение (39) для скорости звука можно получить из уравнения Ньютона (26) и калибровочного условия (28). Действительно, так как (26,28) - это уравнения псевдо непрерывности (уравнения непрерывности со знаком минус), и их функции связаны между собой соотношениями: $p_i = \rho \upsilon_i$, $\sigma_{pj} = \beta^{-1} A_{pj}$, то из уравнения Ньютона (26) следует, что $\sigma_{ij} = -p_i v_j = -\rho \upsilon_i v_j$. Здесь $v_j$ - скорость потока, а $\upsilon_i$ - поле скорости (в гидродинамике Эйлера [1] эти скорости совпадают (4)). Из выражения (28) следует, что $A_{ij} = -c^{-2} \upsilon_i v_j$. Подставляя $A_{ij} = -c^{-2} \upsilon_i v_j$ и $\sigma_{ij} = -\rho \upsilon_i v_j$ в (36), получим скорость звука (39). Выражение (39) можно получить, подставляя $p_i = \rho \upsilon_i$ и $\sigma_{pj} = \beta^{-1} A_{pj}$ в (26) с учетом (28).

Таким образом, уравнения состояния (34,35) фононного поля в сплошной среде, при условии (36) представляют собой уравнения самосогласованного поля (37,38). Выражение (36) учитывает взаимодействие фононного поля со сплошной средой. Калибровочное условие (28) так фиксирует произвол фононного поля, что совместно с уравнениями Ньютона (26) приводит к самосогласованному полю (36). Верно и обратное утверждение: зависимость (36) и уравнения (26) приводят к калибровочному условию (28). В отличие от закона Гука в теории упругости [4], который представляет собой линейное приближение, обобщение закона Гука $\sigma_{pj} = \beta^{-1} A_{pj}$ (36), эквивалентное калибровочному условию (28), фиксирует произвол фононного поля и, как условие, является точным.

Для самосогласованного поля система уравнений состояния лагранжиана (30) разрешима, и все двенадцать компонент фононного поля $A_{pj}$, $\upsilon_p$ удовлетворяют одному и тому же волновому уравнению (37,38), так как $\beta^{-1} = c^2 \rho$ (39). Это уравнение по форме эквивалентно уравнению Клейна-Гордона [9], как линейному неоднородному уравнению Даламбера. Неоднородность уравнений (37,38) обусловлена ненулевой плотностью $\rho$ сплошной среды.

Будем искать решение уравнений состояния (37,38) в виде:

$$A_{pj} = \overline{A}_{pj} \exp(i\vec{q}\vec{X} - i\omega T), \ \upsilon_p = \overline{\upsilon}_p \exp(i\vec{q}\vec{X} - i\omega T), \qquad (40)$$

где $\overline{A}_{pj}$, $\overline{\upsilon}_p$ - не зависят от времени и координат. Тогда из (37,38,40) получим закон дисперсии для фононного поля:

$$\omega = c\sqrt{\vec{q}^{\,2} + c^2 \rho/\gamma} \,. \tag{41}$$

Согласно (41) минимальная частота равна:

$$\omega_0 = c^2\sqrt{\rho/\gamma}\,, \tag{42}$$

а групповая скорость:

$$\frac{\partial \omega}{\partial q} = \frac{c\,|\vec{q}|}{\sqrt{\vec{q}^{\,2} + c^2\rho/\gamma}}\,. \tag{43}$$

Следовательно, скорость волны фононного поля равна групповой скорости (43), а не скорости звука $c$ (39).

Из (42) можно вычислить размерный коэффициент лагранжиана (30): $\gamma = c^4 \rho \omega_0^{-2}$, и выразить $\gamma$ в (30) через $\omega_0$.

Очевидно, что минимальная частота $\omega_0$ находится ниже звукового диапазона 20Гц-20кГц. Положим, что в воздухе $\omega_0 = 3{,}2 \textit{Гц}$. Здесь выбор минимальной частоты условный, и сделан таким для удобства дальнейших вычислений.

Известно, что $c \approx 340\,м/с$, $\rho \approx 1{,}2\,кг/м^3$. Следовательно, для воздуха: $\gamma \approx 1{,}6 \cdot 10^9\,дж \cdot м^2$. Групповая скорость (43), при частоте 20Гц, равна:

$\dfrac{\partial \omega}{\partial q} = \dfrac{c}{\sqrt{1+\dfrac{c^2\rho}{q^2\gamma}}} = \dfrac{c}{\sqrt{1+\dfrac{\omega_0^2}{\omega^2-\omega_0^2}}} = \dfrac{c}{\sqrt{1+1/39}}$. Следовательно, на расстоянии 30 метров запаздывание низких частот будет порядка 1 миллисекунды, в то время как человек способен различать интервалы в 60 миллисекунд. Поэтому, для диапазона 20Гц-20кГц, можно пренебречь дисперсией звука на расстояниях меньше километра.

Нелинейный закон дисперсии (41) обусловлен наличием в лагранжиане (30) механической составляющей $p_i v_i$ (или $\rho v^2/2$). Она дает вклад в левую часть уравнения Даламбера (33) - $p_i$, и делает его неоднородным. Следовательно, левая часть (34,35,37,38) не может быть аннулирована когда $\rho \neq 0$. Случай $\rho = 0$ не соответствует сплошной среде, он приводит к однородному уравнению Даламбера для фононного поля с линейным спектром $\omega = |\vec{q}|c$.

Обратим внимание на то, что механические колебания существуют и для частот меньше $\omega_0$. Для лагранжиана (30) всегда есть решение вида $A_{ij} = \partial u_i/\partial X_j$, $v_i = \partial u_i/\partial T$. В этом случае лагранжиан (23) равен нулю и вариация лагранжиана (30) по $u_i$ приводит к уравнениям Ньютона (26). Подставляя в уравнения (26) $p_i = \rho\, \partial u_i/\partial T$, получим уравнения состояния для механических колебаний в сплошной изотропной среде $\partial \sigma_{ij}/\partial X_j = \rho\, \partial^2 u_i/\partial T^2$ [4], где $\sigma_{ij} = \dfrac{\partial F}{\partial u_{ij}}$, а $F$ потенциал теории упругости (22).

На этом примере видно как система двенадцати уравнений состояния для фононного поля (37,38) трансформируется в три уравнения для вектора смещения. В [4] уравнение Ньютона следует из определения тензора напряжений, и представляют собой уравнения движения. Для уравнений состояния фононного поля уравнения Ньютона представляют собой дифференциальную запись закона сохранения импульса и следуют из трансляционной инвариантности (16,17) лагранжиана (21) или градиентной инвариантности лагранжиана (30). По сути дела в механике сплошной среды исследуются только три уравнения состояния (26) вместо двенадцати уравнений состояния (34,35) для фононного поля, которое обобщает производные вектора смещения.

Заметим, что описание фононного поля в сплошной среде похоже на описание бозона Хиггса в калибровочной теории поля [9,10]. Также как в теории Хиггса:

1) Фононное поле изначально было безмассовым полем взаимодействия и выполняло роль компенсирующего поля в удлиненной производной (5-7) ПП (или волновой функции).

2) Линейная зависимость между сопряженными величинами в сплошной среде $\sigma_{pj} = \beta^{-1} A_{pj}$ и $p_i = \rho \upsilon_i$ приводит к нарушению градиентной инвариантности $A_{ij} \to A_{ij} + \partial u_i / \partial X_j$, $\upsilon_i \to \upsilon_i + \partial u_i / \partial T$ уравнений состояния (32,33).

3) При условии $\rho \neq 0$, фононное поле имеет нелинейный спектр с минимальной частотой (41,42), что делает его «массивным».

В отличие от механизма Хиггса появление массы здесь связано не со спонтанным нарушением калибровочной симметрии, а с приближением сплошной среды, в которой сопряженные величины пропорциональны друг другу. В приближении сплошной среды нет взаимодействия фононного поля и ПП ($a_p = 0$, $k_p = 0$). Можно сказать, что фононное поле материализуется в сплошной среде в прямом и переносном смысле этого слова.

Примером состояния с самосогласованным полем для уравнений Максвелла является сверхпроводимость. Линейная зависимость между сопряженными величинами током $\vec{j}$ и электромагнитным потенциалом $\vec{A}$, приводит к уравнениям Лондонов и энергетической щели в спектре сверхпроводящих электронов [7]. Как известно, уравнения Лондонов описывают эффект Мейснера. Можно полагать, что в стационарном случае будет наблюдаться аналогичный эффект для сопряженных тензоров напряжения и дисторсии в сплошной среде. Однако обсуждение этого вопроса выходит за рамки данной работы.

Целью данной работы является описание фоновой диаграммы Флетчера-Менсона [2]. Теория Хиггса помогает понять, почему компенсирующее фононное поле $A_{pj}$ в сплошной среде с плотностью $\rho$ обладает минимальной частотой, а в воздухе зависит от давления $p$.

## ПЛОТНОСТЬ ЭНЕРГИИ ФОНОННОГО ПОЛЯ И ДИАГРАММА ФЛЕТЧЕРА-МЕНСОНА

Перейдем к исследованию звука в воздухе. Как известно, для воздуха $\sigma_{pj} = -\delta_{pj} p$, здесь $p$ - давление в звуковой волне. Тогда из (36) следует, что давление для самосогласованного поля равно $\delta_{pj} p = -A_{pj} \beta^{-1}$. Это означает, что все диагональные элементы матрицы $A_{pj}$ равны между собой $A_{11} = A_{22} = A_{33} = -\beta p$, а недиагональные равны нулю. Согласно (40) давление представляет собой волну $p = \bar{p} \exp(i\vec{q}\vec{X} - i\omega T)$ со спектром (41), где $\bar{p} = -\bar{A}_{11} \beta^{-1}$. Так как в воздухе $\rho \neq 0$, то согласно (39) $\beta^{-1} \neq 0$, и давление $\bar{p}$ не равно нулю. Следовательно, переход от тензора дисторсии к давлению возможен только в случае $\rho \neq 0$. При этом $p$ - это звуковое давление, которое индуцировано фононным полем $A_{pj}$. Т.е., малое изменение давления $p$ связано с малым изменением дисторсии, а не плотности. Плотность воздуха $\rho$ и сжимаемость $\beta$ здесь константы (37,38). При этом звуковое давление пропорционально дисторсии $p = -c^2 \rho A_{11}$, а не изменению плотности воздуха, см. (64.4) в [1]. Изменение плотности воздуха в звуковой волне описывается здесь как разряжение воздуха выражением $-\rho A_{11}$.

Таким образом, основными параметрами, характеризующими звук в воздухе (2,13), являются скорость и давление, которые имеют волновые решения $\upsilon_p = \bar{\upsilon}_p \exp(i\vec{q}\vec{X} - i\omega T)$, $p = \bar{p} \exp(i\vec{q}\vec{X} - i\omega T)$ с одинаковым спектром (41). При этом уравнения состояния (37,38) – линейные, в отличие от линеаризованных уравнений Эйлера (4) [1], и их волновые решения - точные. Единственное предположение, которое было сделано, – это переход от внешних параметров в (34,35) к самосогласованному полю в сплошной среде. Полученные волновые решения (40) уравнений (37,38), с одинаковым спектром (41) наблюдаемых параметров $p$ и $\upsilon_p$, подтверждает корректность выбора условий (28) и (36).

Подставляя $A_{pj} = -\beta \delta_{pj} p$ в (2,13), получим выражения для напряженностей фононного поля в воздухе:

$$\varepsilon_{pj} = -\frac{\partial \upsilon_p}{\partial X_j} - \delta_{pj} \beta \frac{\partial p}{\partial T} , \qquad (44)$$

$$\rho_{pj} = \beta \, e_{jkp} \frac{\partial p}{\partial X_k} . \qquad (45)$$

В матрице $\varepsilon_{pj}$ (44) временная производная давления содержится только в диагональных элементах. В матрице $\rho_{pj}$ (45) производные давления по координате содержатся только в недиагональных элементах, что соответствует плотности краевых дислокаций. Плотность дислокаций в воздухе представляет собой линейные вихри разряжения, которые пропорциональны градиенту давления. При этом линии дислокаций (линии вихрей), градиент давления и вектор Бюргерса перпендикулярны друг другу согласно (45) (по определению тензора Леви-Чивита). Из (44,45) следует, что **в воздухе звук обусловлен изменением давления и градиентом скорости.**

Подставляя решение (40) в (44,45) получим: $\varepsilon_{pj} = -iq_j \upsilon_p - i\beta \delta_{pj} \omega p$, $\rho_{pj} = i\beta e_{jkp} q_k p$. Здесь целесообразно нормировать $\varepsilon_{pj}$ на скорость $c$ : $\varepsilon'_{pj} = \varepsilon_{pj}/c$ , тогда напряженности $\varepsilon'_{pj}$, $\rho_{pj}$ будут иметь одинаковую размерность $[1/м]$ :

$$\varepsilon'_{pj} = -iq_j \frac{\upsilon_p}{c} - i\delta_{pj} \frac{\omega}{c} \beta p , \; \rho_{pj} = ie_{jkp} q_k \beta p . \qquad (47)$$

Учитывая, что в воздухе $\beta \approx p_{атм}^{-1}$, из (47) следует, что напряженности фононного поля равны относительному изменению давления и скорости в звуковой волне.

Из (30) можно получить выражение для энергии звуковой волны. Первые два члена (30) соответствуют кинетической и потенциальной энергии теории упругости (22). Учитывая, что в воздухе $\sigma_{pj} = -p\delta_{pj}$, то первые два члена лагранжиана (30) приводят к известному выражению для энергии звуковой волны $E = \beta p^2$ (1) [1].

Из второй части лагранжиана (30), получим энергию, обусловленную фононным полем [5,6]:

$$E = \frac{\gamma}{2}(\varepsilon'_{pj}\varepsilon'_{pj} + \rho_{pj}\rho_{pj}) , \qquad (48)$$

которая аналогична плотности энергии электромагнитного поля $\frac{1}{8\pi}(\vec{E}^2 + \vec{B}^2)$. Учитывая, что $\frac{\upsilon}{c} = \beta p$ [1], $q^2 = (\omega^2 - \omega_0^2)/c^2 \approx \omega^2/c^2$ (41) для звукового диапазона частот 20Гц-20кГц при $\omega_0 = 3,2 Гц$, из (42,47,48) получим выражение плотности энергии для продольной звуковой волны:

$$E = 4\beta p^2 \frac{\omega^2}{\omega_0^2} . \qquad (3)$$

Усредняя по времени (3) получим:

$$E = 2\beta\bar{p}^2 \frac{\omega^2}{\omega_0^2}. \qquad (49)$$

Согласно (1,3) усредненная суммарная плотность энергии будет иметь вид:

$$E = \beta\bar{p}^2 \frac{4\omega^2 + \omega_0^2}{2\omega_0^2}.$$

При минимальной частоте $\omega = \omega_0$ плотность механической энергии (1) и фононной энергии (48) соразмерны. Действительно, так как $q^2 = (\omega^2 - \omega_0^2)/c^2$, то при $\omega = \omega_0$ - $q = 0$, и из (48) получим плотность энергии фононов в виде:

$$E(\omega_0) = \frac{3}{4}\beta\bar{p}^2. \qquad (50)$$

Суммарная усредненная энергия при $\omega = \omega_0$ будет равна: $E(\omega_0) = \frac{5}{4}\beta\bar{p}^2$.

Таким образом, в диапазоне частот 20Гц-20кГц энергией (1) можно пренебречь, и исследовать только плотность энергии фононного поля (49). Согласно (49) при изменении частоты от 20Гц до 20кГц плотность энергии изменяется в $10^6$ раз. Поэтому, для графического изображения линий равной плотности энергии на диаграмме давление-частота, перейдем к логарифмической шкале измерений. Логарифмируя (49) по основанию 10 получим: $2\lg\frac{\bar{p}}{\bar{p}_0} = -2\lg\frac{\omega}{\omega_0} + \lg(\frac{E}{2\beta\bar{p}_0^2})$, здесь $\bar{p}_0 = 2\cdot 10^{-5}$ Па - минимальное звуковое давление. Перепишем это выражение в виде:

$10\lg\frac{\bar{p}}{\bar{p}_0} = -10\lg\frac{\omega}{\omega_0} + 5\lg\frac{E}{E_0}$, где $E_0 = 2\beta\bar{p}_0^2$. Обозначим относительное давление, измеряемое в децибелах, как: $L_p = 10\lg\frac{\bar{p}}{\bar{p}_0}$, и исследуем график функции:

$$L_p(\omega, E) = -10\lg\frac{\omega}{\omega_0} + 5\lg\frac{E}{E_0} \qquad (51)$$

при фиксированной плотности энергии $E$.

Так как $\beta \approx 0{,}7\cdot 10^{-5}\,Па^{-1}$ согласно (38), то $E_0 \approx 5{,}6\cdot 10^{-15}\,дж/м^3$. Положим $E_Z = 10^{2Z}\,дж/м^3$, где $Z$ целое число, тогда $L_p(E_Z)$ - линии равной плотности энергии с шагом 10. Для $E = 1\,дж/м^3$ давление в децибелах будет иметь вид: $L_p(\omega) = -10\lg\frac{\omega}{\omega_0} + 71{,}3$. Значение давления для частот $\omega = 20\,Гц$, $\omega = 100\,Гц$, $\omega = 1000\,Гц$ будет соответственно: $L_p(20) = 63{,}3$, $L_p(100) = 56{,}3$, $L_p(1000) = 46{,}3$.

При изменении частоты от 20 Гц $\omega \to \omega_0$, где $\omega_0 = 3{,}2\,Гц$, выражение $L_p(\omega)$ растет не только в связи с уменьшением частоты, но и в связи с тем, что коэффициент в выражении (49) уменьшается с 2 до 3/4 в (50). Такое изменение коэффициента связано с тем, что $q(\omega_0) = 0$ (41) для фононного поля (47). Согласно (1,3,49,50) плотность механической энергии и энергии фононного поля пропорциональна квадрату давления при любой частоте, поэтому интервал между линиями равной плотности энергии не меняется с частотой.

На фоновой диаграмме Флетчера-Менсона также наблюдается изменение угла наклона линий изофон в районе низких частот. Но при этом наблюдается и уменьшение интервала между линиями изофон на рис. 1. Это сужение обусловлено субъективными факторами, оно отражает факт существования минимальной частоты фононного поля.

На участке 20Гц-1кГц линии равной плотности энергии (51) имеют вид отрицательного логарифма и, по форме, совпадают с изофонами. Такое поведение объясняется квадратичной зависимостью плотности энергии от частоты в этом диапазоне (49).

Выше 1кГц при увеличении частоты давление изофон сначала стабильно, а затем начинает расти рис. 1. Для линий с одинаковой плотностью энергии давление продолжает плавно уменьшатся (51) по логарифмическому закону. Представляется, что это отличие связано с более интенсивным поглощением звука на высоких частотах.

Как известно [1], коэффициент поглощения звука $\alpha$ пропорционален квадрату частоты. Кроме этого коэффициент поглощения звука зависит от макроскопических параметров воздуха: температуры, влажности, плотности и т.д. Учет поглощения звука $p = pe^{-\alpha X}$ для функции $L_p$ (51) приводит к дополнительной квадратичной зависимости от частоты:

$$L_p(\omega, E) = \chi\omega^2 - 10\lg\frac{\omega}{\omega_0} + 5\lg\frac{E}{E_0}, \tag{52}$$

где $\chi\omega^2 = \alpha d \cdot 10\lg e$, здесь $d$ - расстояние до источника звука, $\chi$ - коэффициент, не зависящий от частоты $\omega$.

Предположим, что для 2кГц поглощение $\chi\omega^2$ равно 1дБ, тогда для 4кГц поглощение будет равно 4 дБ, 8кГц -16дБ, 10кГц – 25дБ. При этом соответствующее увеличение частоты приводит к уменьшению $L_p$ за счет $-10\lg\frac{\omega}{\omega_0}$. Так, для 4кГц $L_p$ уменьшится за счет $-10\lg\frac{\omega}{\omega_0}$ по сравнению со значением при 2кГц на 3дБ, 8кГц - 6дБ, 10 кГц - 7дБ, соответственно. Складывая эти изменения, получим, что минимум зависимости $L_p$ от частоты лежит между 2кГц и 4кГц, выше 4кГц функция $L_p$ изменяется подобно параболе, что соответствует рис. 1.

Как известно [1], описанное поглощение звука на высоких частотах происходит на расстояниях более 200 метров от источника звука. Следовательно, субъективное восприятие громкости звука отражает факт затухания звуковых волн на частотах выше 2кГц так, как если бы человек находился на расстоянии более 200 метров от источника звука.

При извлечении звука и его усилении не должны учитываться субъективные факторы, а также поглощение звука на расстоянии. Построим линии равной плотности энергии в логарифмических координатах. Введем, по аналогии с логарифмической шкалой децибелов для давления, логарифмические выражения характеризующие частоту $L_\omega = 10\lg\frac{\omega}{\omega_0}$ и энергию $L_E = 10\lg\frac{E}{E_0}$. Тогда линии равной плотности энергии (51) в координатах $(L_\omega, L_p)$ в диапазоне 20Гц-20кГц имеют вид пучка параллельных прямых линий с тангенсом угла $tg\phi = -1$ (рис. 2):

$$L_p = -L_\omega + \frac{1}{2}L_E. \tag{53}$$

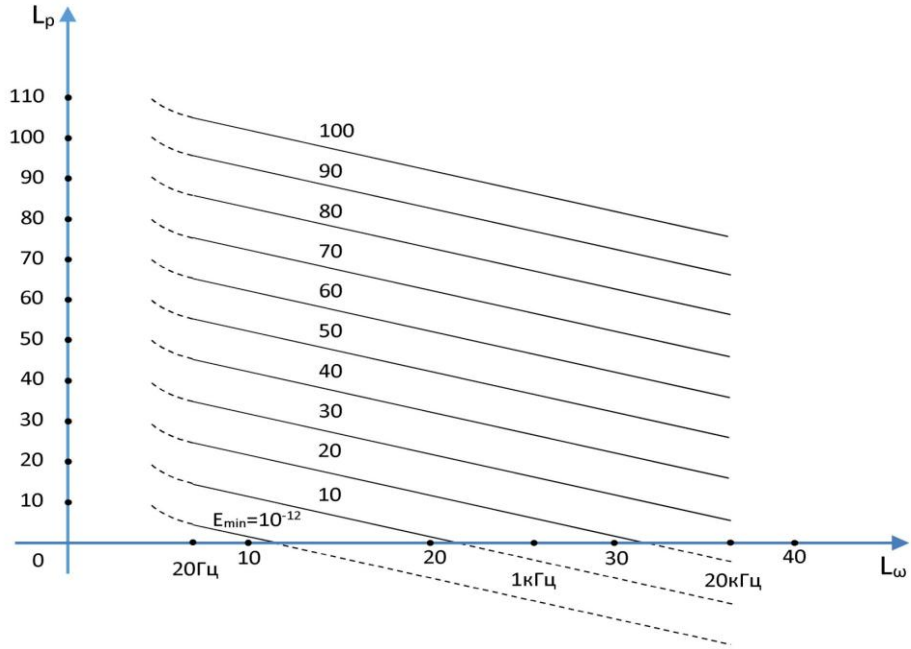

Рис. 2. Линии равной плотности энергии $L_N$ в логарифмических координатах $L_p = 10\lg\dfrac{\overline{p}}{\overline{p}_0}$, где $\overline{p}_0 = 2\cdot 10^{-5} Па$; $L_\omega = 10\lg\dfrac{\omega}{\omega_0}$, где $\omega_0 = 3{,}2 Гц$. Пороговое значение плотности энергии $E_{\min} = 10^{-12} дж/м^3$ - соответствует нулевому уровню $L_0 = 0$. Последующая нумерация уровней определяется по формуле $L_N = 5\lg\dfrac{E_N}{E_{\min}}$, где $E_N = 10^{2N} E_{\min}$.

Ниже 20Гц, при $\omega \to \omega_0$, параллельные прямые переходят в кривые с увеличением величины тангенса угла наклона рис. 2. Это связано с нелинейной зависимостью (41) и наличием механической энергии (1), которая не зависит явно от частоты. Интервал между линиями равной плотности энергии не меняется, так как $\Delta L_p = \dfrac{1}{2}\Delta L_E$ при любой фиксированной частоте (3,49,50).

Из выражения (53) следует, что минимальная плотность энергии (порог слышимости) зависит от частоты, и задается соотношением: $L_E = 2L_\omega$, когда $L_p = 0$. Таким образом, с ростом частоты растет плотность энергии, которая соответствует пороговому давлению $p_0$. При низких частотах порог чувствительности равен $E_{\min} \approx 10^{-12} дж/м^3$, так как $L_p(20) = 63{,}3$, $L_p(100) = 56{,}3$ для $E = 1 дж/м^3$. Учитывая, что пороговая интенсивность звуковой волны равна $E_{\min} \approx 10^{-12} дж/м^3$, построим линии равной плотности энергии. В этом случае нулевой изофоне будет соответствовать линия равной плотности энергии $E_{\min}$: $L_p = -L_\omega + 11{,}25$ рис. 2. Тогда линии равной плотности энергии получим подстановкой $E_N = 10^{2N} E_{\min}$ в (53).

Из рис. 2 видно, что условный выбор нулевой частоты $\omega_0 = 3{,}2 Гц$ определяет шкалу $L_\omega$. При уменьшении нулевой частоты реальные частоты будут смещаться направо, а при увеличении

$\omega_0$ - налево. Из рис. 2 следует, что шкала изофон рис. 1 коррелирует с со шкалой равной плотности энергии.

Таким образом, диаграмма равной плотности энергии в звуковом диапазоне 20Гц-20кГц представляет собой прямые линии в логарифмических координатах рис. 2. Практическое применение диаграммы рис. 2 возможно при усилении звука. Она показывает, как надо усиливать звуковой сигнал, чтобы объективно восстановить звуковую энергию. Здесь мы исходим из того, что в природе нет усиления звука, а есть ослабление звука вследствие его рассеяния и поглощения. Поэтому усиление звука должно основываться на объективных показателях, которым является плотность энергии звуковой волны рис. 2, а не на субъективном восприятии громкости звука рис. 1.

**ЗАКЛЮЧЕНИЕ**

В работе было предложено описание звука в виде напряженностей $\varepsilon_{pj}$, $\rho_{pj}$ (2,13) фононного поля $A_{pj}$, $\upsilon_p$ (4-тензора дисторсии). Для напряженностей фононного поля $\varepsilon_{pj}$, $\rho_{pj}$ получены уравнения состояния (32,33), аналогичные уравнениям Максвелла, которые сводятся к неоднородным уравнениям Даламбера (34,35). Показано, что эти уравнения приводят к волновым решениям (40) для самосогласованного поля в сплошной среде: $p_i = \rho \upsilon_i$, $\sigma_{pj} = \beta^{-1} A_{pj}$ (37,38), со спектром (41) и минимальной частотой (42). В воздухе напряженности $\varepsilon_{pj}$, $\rho_{pj}$ являются функциями давления и скорости (44,45). Плотность энергии (49) звуковой волны пропорциональна квадрату частоты и давления в диапазоне 20Гц-20кГц. Линии равной плотности энергии (51) имеют вид отрицательного логарифма и, по форме, совпадают с изофонами диаграммы Флетчера-Менсона [2,3] в среднем звуковом диапазоне рис. 1.

Субъективное восприятие громкости звука на диаграмме Флетчера-Менсона проявляется на низких и высоких частотах:
1) на низких частотах оно отражает факт существования минимальной частоты, которая находится ниже звукового диапазона;
2) на высоких частотах оно отражает квадратичный закон поглощения звука с расстоянием в зависимости от частоты (52). Согласно диаграмме рис. 1 субъективное восприятие сводится к усилению давления на высоких частотах по квадратичному закону. Это усиление эквивалентно поглощению высоких частот, которое возникает на расстоянии более 200 метров от источника звука.

В связи с тем, что звук в природе распространяется независимо от его восприятия, предложено при усилении звукового сигнала пользоваться диаграммой рис.2 (53), которая не зависит от субъективных факторов.

Может сложиться впечатление, что волновые решения для давления $p$ и скорости $\upsilon_p$ в воздухе, которые были получены из уравнений состояния (37,38) для фононного поля, аналогичны решениям линеаризованных уравнений Эйлера [1]. Однако это не так.

Для воздуха совпали параметры $\upsilon_i$ и $p$, которые характеризуют напряженности фононного поля (44,45), с параметрами (4), которые используются в уравнениях Эйлера [1]. Первые из них описывают звук (47) с нелинейным спектром (41), вторые – механические колебания сплошной среды [1] с линейным спектром $\omega = |\vec{q}| c$. Чтобы понять, какая теория описывает звук, надо сравнить плотность энергии звуковой волны в обоих случаях с экспериментом.

Волновые решения линеаризованных уравнений Эйлера были получены исходя из предположения, что параметры $\upsilon_p$ и $p$ - малы в звуковой волне, поэтому нелинейными членами в (4) решили пренебречь [1]. Однако, полученные решения для давления и скорости, см. (64.6, 64.18) в [1]: $\upsilon_p = \partial \varphi / \partial x_p$, $p = -\rho \partial \varphi / \partial t$ - пропорциональны частоте, так как функция $\varphi = \text{Re}(\varphi_0 \exp(-i\omega(t - x/c)))$ удовлетворяет однородному уравнению Даламбера. Поэтому такие периодические решения $\upsilon_p$ и $p$ не могут считаться малыми для высоких частот. Это противоречит предположению, согласно которому волновые решения здесь были получены. Более того, линейная зависимость величины давления от частоты $p = \omega\rho \, \mathrm{Re}(i\varphi_0 \exp(-i\omega(t-x/c)))$ прямо противоречит диаграмме Флетчера-Менсона (рис. 1).

Как было показано выше, механическое представление о звуке [1] не может в принципе объяснить (1) фоновую диаграмму Флетчера-Менсона – зависимость давления от частоты в изофонах. В воздухе $A_{pj} = -\beta\delta_{pj}p$, так как в сплошной среде $A_{pj} = \beta\sigma_{pj}$ (36). Поэтому при описании звука в воздухе, тензор дисторсии можно заменить звуковым давлением в (2,13). При этом плотность воздуха $\rho$ и сжимаемость $\beta$ предполагаются константами, а изменение плотности воздуха в звуковой волне описывается выражением - $-\rho A_{11}$. Напряженности фононного поля $\varepsilon_{pj}$ и $\rho_{pj}$ (47), для волновых решений $\upsilon_p$ и $p$, дают основной вклад в плотность энергии звуковой волны (48,49) - пропорциональный квадрату частоты и давления, что соответствует диаграмме Флетчера-Менсона рис. 1.

Как известно, уравнения Эйлера представляют собой суперпозицию сил, действующих на единичный объем. Для звуковой волны в воздухе его нужно обобщить или дополнить силами, действующими на импульс в фононном поле (14,15). Так как в воздухе $A_{pj} = -\beta\delta_{pj}p$, то центрально-симметричная сила (14) будет равна $f_j^C = -p_p \partial\upsilon_p/\partial X_j - p_j\beta\partial p/\partial T$, а сила Пича-Келлера (15) будет иметь вид $f_j^A = 2\beta p \partial p/\partial X_j$. Таким образом, в уравнение Эйлера добавится акустическая сила обусловленная изменением давления со временем: $-p_j\beta\partial p/\partial T$, и сила Пича-Келлера, которая в сумме с градиентом давления будет иметь вид: $-(1-2\beta p)\partial p/\partial X_j$.

Нелинейный закон дисперсии (41) и минимальная частота (42) фононного поля обусловлены существованием звука в сплошной изотропной среде в виде самосогласованного поля: $p_i = \rho\upsilon_i$ и $\sigma_{pj} = \rho c^2 A_{pj}$ (39). Эти выражения описывают взаимодействие фононного поля со сплошной средой. Переход к самосогласованному полю сопровождается нарушением градиентной инвариантности уравнений состояния (32,33) и (37,38). Этот переход похож на механизм Хиггса появления массивного бозона [10]. Аналогия с механизмом Хиггса для фононного поля в сплошной среде прослеживается в связи с тем, что фононное поле, как компенсирующее поле ПП, изначально было безмассовым полем взаимодействия (5-7), (23). Представляется, что фононное поле существует в сплошной изотропной среде с плотностью $\rho$ в виде самосогласованного поля, по определению сплошной среды.

Минимальная частота звуковых колебаний (42) является следствием наличия в энергии двух слагаемых: механической энергии (1) и энергии фононного поля (3,48). Вклад соответствующих слагаемых в лагранжиан (30) характеризуется размерными множителями $\rho$ и $\gamma$. Коэффициент $\rho\gamma^{-1}$ входит в левую часть уравнений состояния (37,38), так как $\beta^{-1} = c^2\rho$ (39), и задает неоднородность уравнений Даламбера. Про параметры $\rho$ и $\gamma$ можно сказать, что они характеризуют механические и фононные степени свободы, так как условие $\rho = 0$ - аннулирует механическую энергию, а $\gamma = 0$ - аннулирует фононную энергию. Квадрат минимальной частоты пропорционален $\rho\gamma^{-1}$ (42), и задает границу, ниже которой не существует фононных волн.

**СПИСОК ЛИТЕРАТУРЫ**